\newcommand*{\ATLASLATEXPATH}{}
\documentclass[cernpreprint,texlive=2011,txfonts,UKenglish]{\ATLASLATEXPATH atlasdoc}

\pdfoutput=1
\usepackage{\ATLASLATEXPATH atlaspackage}
\usepackage{\ATLASLATEXPATH atlasbiblatex}

\usepackage{\ATLASLATEXPATH atlascontribute}

\usepackage{\ATLASLATEXPATH atlasphysics}
\graphicspath{}

\usepackage{excdilept_paper-defs}
\usepackage{graphicx}
\usepackage{makecell}
\usepackage{overpic}
\usepackage{lineno}

\AtlasTitle{Measurement of exclusive \gmgmll production in proton--proton collisions at $\sqrt{s} = 7~\TeV$ with the ATLAS detector}

\author{The ATLAS Collaboration}

 \AtlasJournal{Phys.\ Lett.\ B.}

\AtlasAbstract{
This Letter reports a measurement of the exclusive \gmgmll ($\ell=e,~\mu$) cross-section in proton--proton collisions 
at a centre-of-mass energy of $7~\TeV$ by the ATLAS experiment at the LHC, based on an integrated luminosity of $4.6~\ifb$.
For the electron or muon pairs satisfying exclusive selection criteria, a fit to the dilepton acoplanarity distribution is used 
to extract the fiducial cross-sections.
The cross-section in the electron channel is determined to be 
$\sigma_{\gmgmee}^{\mathrm{excl.}} = 0.428~\pm~0.035~(\mathrm{stat.})~\pm~0.018~(\mathrm{syst.})~\mathrm{pb}$ for a phase-space region with invariant mass of the electron pairs greater than $24~\GeV$, in which both electrons have transverse momentum $\pT>12~\GeV$ and pseudorapidity $|\eta|<2.4$.
For muon pairs with invariant mass greater than $20~\GeV$, muon transverse momentum $\pT>10~\GeV$ and pseudorapidity $|\eta|<2.4$, the cross-section is determined to be $\sigma_{\gmgmmumu}^{\mathrm{excl.}} = 0.628~\pm~0.032~(\mathrm{stat.}) ~\pm~0.021~(\mathrm{syst.})~\mathrm{pb}$.
When proton absorptive effects due to the finite size of the proton are taken into account in the theory calculation,
the measured cross-sections are found to be consistent with the theory prediction.
}

\PreprintIdNumber{CERN-PH-EP-2015-134}

\begin{document}

\maketitle

\tableofcontents

% List of contributors - print here or after the Bibliography
%\AtlasPrintContribute{0.3}
%\clearpage

\section{Introduction}
\label{sec:intro}
A considerable fraction of proton--proton ($pp$) collisions at high energies involve reactions mediated by photons. 
This fraction is dominated by elastic scattering, with a single photon exchange. 
Quasi-real photons can also be emitted by both protons, with a variety of final states produced. 
In these processes the $pp$ collision can be then considered as a photon--photon ($\gamma\gamma$) collision.
At the LHC, these reactions can be studied at energies well beyond the electroweak energy scale~\cite{Chatrchyan:2013foa}.
The cross-section of the $pp (\gamma \gamma) \rightarrow \ell^+ \ell^- X$ process has been predicted to increase with energy~\cite{Carimalo:1978bu} and constitutes a non-negligible background to Drell--Yan (DY) reactions~\cite{Aad:2013iua}.

The exclusive two-photon production of lepton pairs
($pp (\gamma\gamma) \rightarrow \ell^+\ell^- pp$, referred to as exclusive $\gmgmll$) can be calculated in the framework of quantum electrodynamics (QED)~\cite{Chen:1973mv, Budnev:1973tz}, within uncertainties of less than 2\% associated with the proton elastic form-factors.
Exclusive dilepton events have a clean signature that helps discriminate them from background: there are only two identified
muons or electrons, without any other activity in the central
detectors, and the leptons are back-to-back in azimuthal angle.
Furthermore, due to the very small photon virtualities involved, the incident protons are scattered at almost zero-degree angles.
Consequently, the measurement of exclusive $\gmgmll$~reactions was proposed for precise absolute luminosity measurement at hadron colliders~\cite{Budnev:1973tz, Khoze:2000db, Shamov:2002yi, Krasny:2006xg}.
However, this process requires significant corrections (of the order of 20\%) due to additional interactions between the elastically scattered protons~\cite{Schafer:2007mm, Dyndal:2014yea}.

At hadron colliders exclusive $\gmgmll$~events have been observed
in $ep$ collisions at HERA \cite{Aktas200428}, in $p\bar{p}$ collisions at the Tevatron \cite{Abulencia:2006nb, Aaltonen:2009cj, Aaltonen:2009kg} and in nucleus--nucleus collisions
at RHIC~\cite{Adams:2004rz, Afanasiev:2009hy} and the LHC \cite{Abbas:2013oua}.
The exclusive two-photon production of lepton pairs in $pp$ collisions at the LHC was studied recently by the CMS collaboration~\cite{Chatrchyan:2011ci, Chatrchyan:2012tv}.

This Letter reports a measurement of exclusive dilepton production in $pp$~collisions at $\sqrt{s} = 7~\TeV$. 
The measurement of exclusive dilepton production cross-section is compared to the QED-based prediction with and without proton absorptive corrections.

\section{The ATLAS detector}
The ATLAS experiment~\cite{Aad:2008zzm} at the LHC is a multi-purpose particle detector with a forward-backward symmetric cylindrical geometry and nearly 4$\pi$ coverage in solid angle.\footnote{ATLAS uses a right-handed coordinate system with its origin at the nominal interaction point in the centre of the detector and the $z$-axis coinciding with the axis of the beam pipe. The $x$-axis points from the interaction point to the centre of the LHC ring, and the $y$-axis points upward.  The pseudorapidity is defined in terms of the polar angle $\theta$ as $\eta=-\ln \tan(\theta/2)$, and $\phi$ is the azimuthal angle around the beam pipe with respect to the $x$-axis. 
The angular distance is defined as $\Delta R=\sqrt{(\Delta \eta)^2+(\Delta \phi)^2}$. 
The transverse momentum is defined relative to the beam axis.}
It consists of inner tracking devices surrounded by a superconducting solenoid, electromagnetic and hadronic calorimeters, and a muon spectrometer. 
The inner detector (ID) provides charged-particle tracking in the pseudorapidity region $|\eta|<2.5$ and vertex reconstruction. 
It comprises a silicon pixel detector, a silicon microstrip tracker, and a straw-tube transition radiation tracker.
The ID is surrounded by a solenoid that produces a 2 T axial magnetic field. 
Lead/liquid-argon (LAr) sampling calorimeters provide electromagnetic (EM) energy measurements with high granularity. 
A hadron (iron/scintillator-tile) calorimeter covers the central pseudorapidity range $|\eta|<1.7$.
The end-cap and forward regions are instrumented with LAr calorimeters for both the EM and hadronic energy measurements up to $|\eta|=4.9$. 
The muon spectrometer (MS) is operated in a magnetic field provided by air-core superconducting toroids and includes tracking chambers
for precise muon momentum measurements up to $|\eta|=2.7$ and trigger chambers covering the range $|\eta|<2.4$.
 
A three-level trigger system is used to select interesting events. 
The first level is implemented in custom electronics and is followed by two software-based trigger levels, referred to collectively as the High-Level Trigger.

\section{Theoretical background and event simulation}
\label{sec:simulation}

Calculations of the cross-section for exclusive two-photon production of lepton pairs in $pp$ collisions are based on the
Equivalent Photon Approximation (EPA)~\cite{Chen:1973mv,Budnev:1973tz,Weizsacker:1934sx,ww, Terazawa:1973tb, Budnev:1974de}. 
The EPA relies on the property that the EM field of a charged particle, here a proton, moving at high velocity becomes more and more transverse with respect to the direction of propagation.
As a consequence, an observer in the laboratory frame cannot distinguish between the EM field of a relativistic proton and the transverse component of the EM field associated with equivalent photons.
Therefore, using the EPA, the cross-section for the reaction above can be written as

\noindent
$$
\sigma^\mathrm{EPA}_{pp (\gamma \gamma) \rightarrow \ell^+ \ell^- pp} = \iint P(x_1) \, P(x_2) \, \sigma_{\gamma \gamma \rightarrow \ell^+ \ell^-}(\Mll^2) \, \mathrm{d}x_1 \, \mathrm{d}x_2~,
$$
\noindent
where $P(x_1)$ and $P(x_2)$ are the equivalent photon spectra for the  protons, $x_1$~and $x_2$~are the fractions of the proton energy carried away by the emitted photons and $\Mll$~is the invariant mass of the lepton pair. 
These variables are related by $\Mll^2/s=x_1x_2$~where $s$~is the $pp$~centre-of-mass energy squared.  
The symbol $\sigma_{\gamma \gamma \rightarrow \ell^+ \ell^-}$~refers to the cross-section for the QED sub-process.
As discussed previously, the photons are quasi-real, which means that their virtuality $Q^2$ is very small compared to $\Mll^2$.
In this kinematic region the EPA gives the same predictions as full leading-order (LO) QED calculations~\cite{Chen:1973mv,Budnev:1973tz}.

In the reaction $pp (\gamma \gamma) \rightarrow \ell^+ \ell^- X$ the protons scattering can be: elastic, $X=pp$; single-dissociative, $X=pX'$; or double-dissociative, $X=X'X''$
(the symbols $X'$,~$X''$~denote any additional final state produced in the event).
Unless both outgoing protons are detected, the proton dissociative events form an irreducible background to the fully elastic production.

Such photon-induced reactions, in particular exclusive $\gmgmll$ production, require significant corrections due to proton absorptive effects.
These effects are mainly related to $pp$ strong-interaction exchanges that accompany the two-photon interaction and that lead to the production of additional hadrons in the final state.
Recent phenomenological studies suggest that the exclusive $\gmgmll$ cross-section is suppressed by a factor that depends on the mass and rapidity of the system produced~\cite{Dyndal:2014yea}. 
For the kinematic range relevant for this measurement the suppression factor is about 20\%.
This factor includes both the strong $pp$ absorptive correction ($\sim$8\% suppression) and the photon--proton ($\gamma p$) coherence condition ($b_{\gamma p} > r_p$, where $b_{\gamma p}$ is the $\gamma p$ impact parameter and $r_p$ the transverse size of the proton).

Simulated event samples are generated in order to estimate the background and to correct the signal yields for detector effects.
The signal event samples for exclusive \gmgmll~production are generated using the \herwigpp~2.6.3~\cite{Bahr:2008pv} Monte Carlo (MC) event generator, which implements the EPA formalism in $pp$ collisions.
The dominant background, photon-induced single-dissociative dilepton production, is simulated using \lpair~4.0~\cite{Vermaseren:1982cz} with the  Brasse~\cite{Brasse:1976bf} and Suri--Yennie~\cite{Suri1972243} structure functions for proton dissociation. 
For photon virtualities $Q^2 < 5~\GeV^2$ and masses of the dissociating system, $m_N<2~\GeV$, low-multiplicity states from the production and decays of $\Delta$ resonances are usually created. 
For higher $Q^2$ or $m_N$, the system decays to a variety of resonances, which produce a large number of forward particles.
The \lpair~package is interfaced to \jetset~7.408~\cite{Sjostrand:1993yb}, where the \lund~\cite{Andersson:1983ia} fragmentation model is implemented.
The \herwigpp~and \lpair~generators do not include any corrections to account for proton absorptive effects.

For double-dissociative reactions, \pythia~8.175~\cite{Sjostrand:2007gs} is used with the NNPDF2.3QED~\cite{Ball:2013hta} parton distribution functions (PDF).
The NNPDF2.3QED set uses LO QED and next-to-next-to-leading-order (NNLO) QCD perturbative calculations to construct the photon PDF, starting from the initial scale $Q^2_0 = 2~\GeV^2$. 
Depending on the multiplicity of the dissociating system, the default \pythia 8~string or mini-string fragmentation model is used for proton dissociation.
The absorptive effects in double-dissociative MC events are taken into account using the default multi-parton interactions model in~\pythia 8~\cite{Corke:2009tk}.

The \powheg~1.0~\cite{Nason:2004rx, Frixione:2007vw, Alioli:2008gx} MC generator is used with the CT10~\cite{Lai:2010vv} PDF to generate both the DY~\Zstee~and~\Zstmumu~events. 
It is interfaced with \pythia~6.425~\cite{Sjostrand:2006za} using the CTEQ6L1~\cite{Pumplin:2002vw} PDF set and the AUET2B~\cite{ATLAS:2011zja} values of the tunable parameters to simulate the parton shower and the underlying event (UE).
These samples are referred to as \powhegpythia.
The DY \Zsttautau~process is generated using \pythia~6.425 together with the MRST LO*~\cite{Sherstnev:2007nd} PDF. 
The transverse momentum of lepton pairs in \powhegpythia~samples is reweighted to a \resbos~\cite{Landry:2002ix} prediction, which is found to yield good agreement with the transverse momentum distribution of $Z$ bosons observed in data~\cite{Aad:2012wfa, Aad:2014xaa}.
The production of top-quark pair (\ttbar)~events is modelled using
\mcatnlo~3.42~\cite{Frixione:2002ik, Frixione:2003ei} and diboson (\WW, \WZ, \ZZ) processes are simulated using \herwig~6.520~\cite{Corcella:2002jc}.
The event generators used to model \Zg, \ttbar~and diboson reactions are interfaced to \photos~3.0~\cite{Golonka:2005pn} to simulate QED final-state radiation (FSR) corrections.

Multiple interactions per bunch crossing (pile-up) are accounted for by overlaying simulated minimum-bias events, generated with \pythia~6.425 using the AUET2B tune and CTEQ6L1 PDF, and reweighting the distribution of the average number of interactions per bunch crossing in MC simulation to that observed in data.
Furthermore, the simulated samples are weighted such that the $z$-position distribution of reconstructed $pp$ interaction vertices matches the distribution observed in data.
The ATLAS detector response is modelled using the GEANT4 toolkit~\cite{Agostinelli2003250, 2010EPJC} and the same event reconstruction as that used for data is performed.

\section{Event reconstruction, preselection and background estimation}
\label{sec:reco}
The data used in this analysis were collected during the 2011 LHC $pp$ run at a centre-of-mass energy of $\sqrt{s} = 7~\TeV$. After application of data-quality requirements, the total integrated luminosity is $4.6~\ifb$ with an uncertainty of 1.8\% \cite{Aad:2013ucp}.
Events from these $pp$ collisions are selected by requiring at least one collision vertex with at least two charged-particle tracks with $\pT > 400~\MeV$.
Events are then required to have at least two lepton candidates (electrons or muons), as defined below.

Events in the electron channel were selected online by requiring a single-electron or di-electron trigger.
For the single-electron trigger, the transverse momentum threshold was increased during data-taking from 20~\GeV~to 22~\GeV~in response to the increased LHC instantaneous luminosity. 
The di-electron trigger required a minimum transverse momentum of 12~\GeV~ for each electron candidate.
Electron candidates are reconstructed from energy deposits in the calorimeter matched to ID tracks.
Electron reconstruction uses track refitting with a Gaussian-sum filter to be less sensitive to bremsstrahlung losses and improve the estimates of the electron track parameters \cite{Fruhwirth19971, ATLAS:2012dma}.
The electrons are required to have a transverse momentum $\pT^e>12~\GeV$ and pseudorapidity $|\eta^e|<2.4$ with the calorimeter barrel/end-cap transition region $1.37<|\eta^e|<1.52$ excluded.
Electron candidates are required to meet ``medium'' identification criteria based on shower shape and track-quality variables \cite{ref29}.

Events in the muon channel were selected online by a single-muon or di-muon trigger, with a transverse momentum threshold of 18~\GeV~or 10~\GeV, respectively. 
Muon candidates are identified by matching complete tracks in the MS to tracks in the ID \cite{Aad:2014rra}, and are required to have $\pT^{\mu}>10~\GeV$ and $|\eta^{\mu}|<2.4$.
Only isolated muons are selected by requiring the scalar sum of the
\pT~of the tracks with $\pT>1~\GeV$ in a $\Delta R=0.2$ cone around the muon to be less than 10\% of the muon~\pT.

Di-electron (di-muon) events are selected by requiring two oppositely charged same-flavour leptons with an invariant mass $\Mee>24~\GeV$ for the electron channel and $\Mmm>20~\GeV$ for the muon channel.
After these preselection requirements $1.57\times 10^6$~di-electron and $2.42\times 10^6$~di-muon candidate events are found in the data.

The background to the exclusive signal includes contributions from single- and double-proton dissociative \gmgmll~production, as well as \Zg,
diboson, \ttbar~and multi-jet production.
The contribution from $\gamma\gamma \rightarrow W^+W^-$ and $\gamma\gamma  \rightarrow \tau^+\tau^-$ processes is considered negligible.
Single- and double-dissociative background contributions are estimated using MC simulations.
The electroweak (\Zg, diboson) and top-quark pair background contributions are also estimated from simulations 
and normalised to the respective inclusive cross-sections calculated at high orders in perturbative QCD (pQCD), as in Ref.~\cite{Aad:2014qja}.
Scale factors are applied to the simulated samples to correct for the small differences from data in the trigger, reconstruction and
identification efficiencies for electrons and muons~\cite{ref29, Aad:2014rra, Aad:2014qja}.
MC events are also corrected to take into account differences from data in lepton energy, momentum scale and resolution~\cite{Aad:2014rra, ref27}.

The multi-jet background is determined using data-driven methods, similarly to Refs.~\cite{Aad:2013ysa, Aad:2014xaa}.
For the \ee~channel, the multi-jet sample is obtained by applying the full nominal preselection but requiring the electron candidates to not satisfy the medium identification criteria.
For the \mumu~channel, it is extracted using same-charge muon pairs that satisfy the remaining preselection criteria.
The normalisation of the multi-jet background is determined by fitting the  invariant mass spectrum of the electron (muon) pair in the data to a sum of expected contributions, including MC predictions of the signal and the other backgrounds.

\section{Exclusive event selection and signal extraction}
\label{sec:cross-sec}
In order to select exclusive \gmgmll~candidates, a veto on additional charged-particle track activity is applied. 
This exclusivity veto requires that no additional charged-particle tracks with $\pT>400~\MeV$~be associated with the dilepton vertex, and that no additional tracks or vertices be found within a 3~mm longitudinal isolation distance, $\Delta z_\mathrm{vtx}^\mathrm{iso}$, from the dilepton vertex.
These conditions are primarily motivated by the rejection of the \Zg~and multi-jet events, which typically
have many tracks originating from the same vertex.

\begin{figure}[t!]
\begin{center}
  \includegraphics[width=0.49\textwidth]{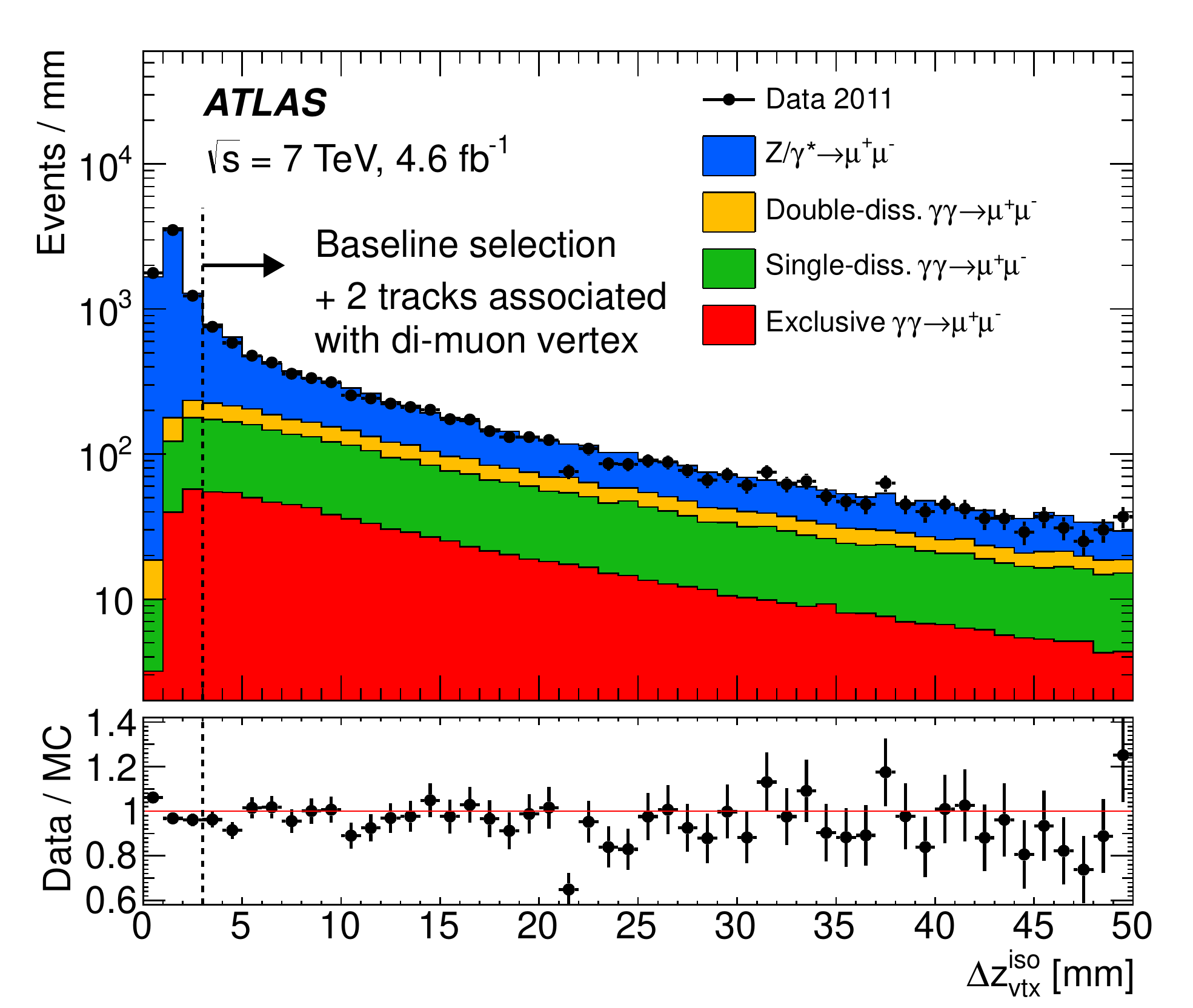}\put(-110,-8){{\footnotesize(a)}}
  \vskip 0pt
  \includegraphics[width=0.49\textwidth]{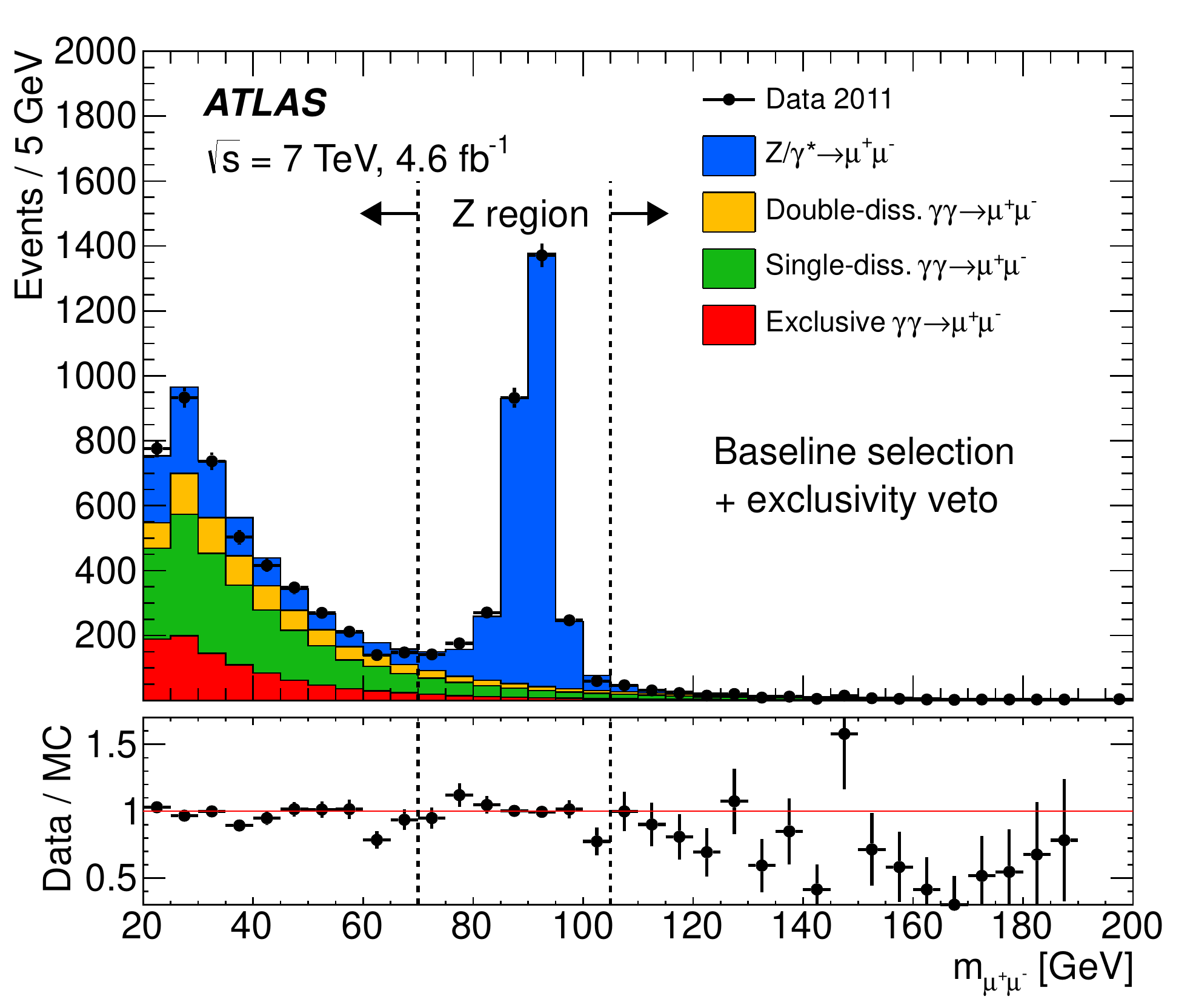}\put(-110,-8){{\footnotesize(b)}}
  \includegraphics[width=0.49\textwidth]{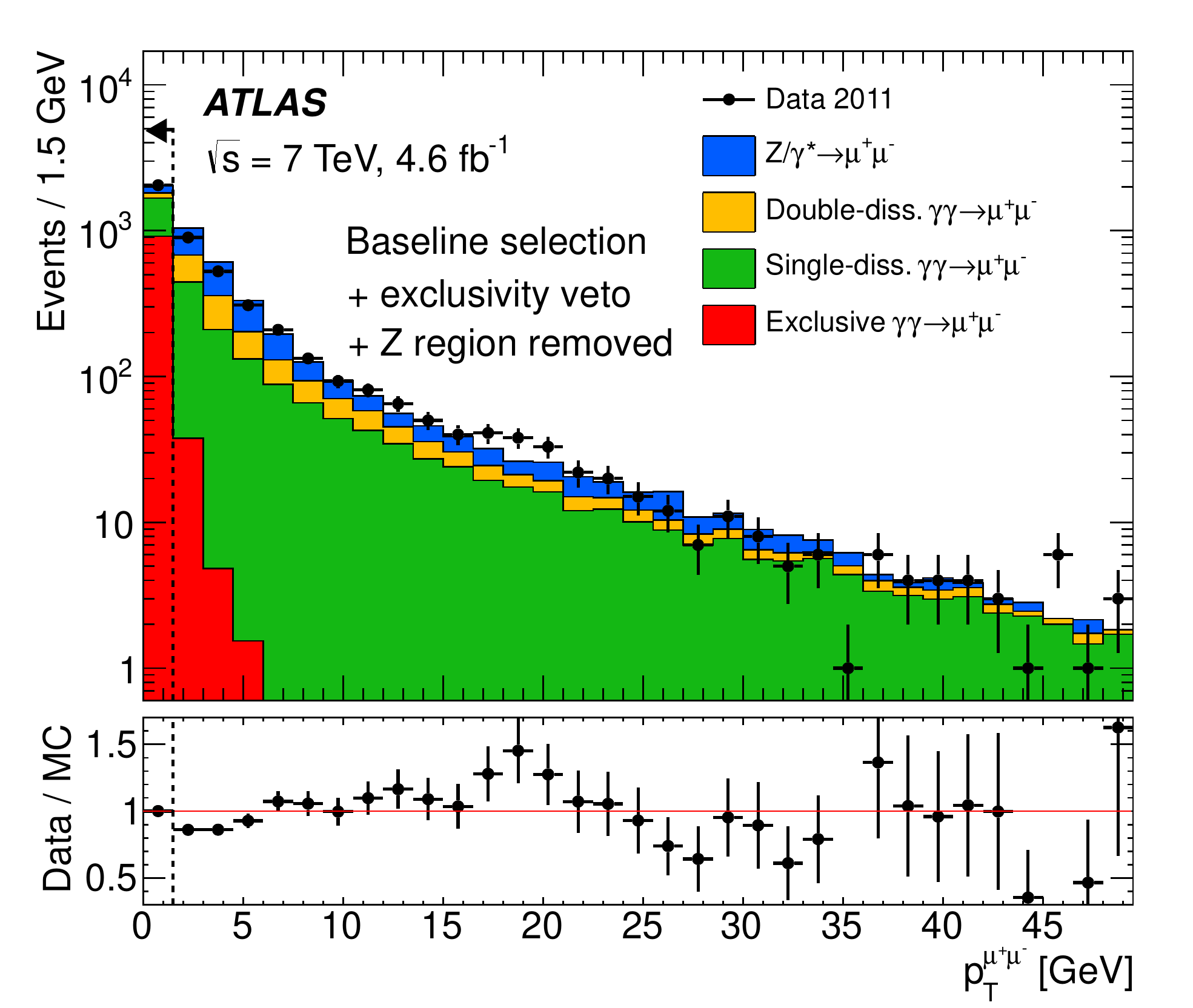} \put(-110,-8){{\footnotesize(c)}}
\end{center}
  \vskip -15pt
\caption{
Illustration of exclusive event selection in the muon channel (see text). 
(a) Longitudinal distance between the di-muon vertex and any other tracks or vertices, (b) di-muon invariant mass, and (c) transverse momentum of
the di-muon system, after application of subsequent selection criteria (indicated by the dashed lines). Data are shown as points with statistical error bars, while the histograms represent the expected signal and background levels, corrected using the scale factors described in the text.
\label{fig:selection}
}
\end{figure}
The charged-particle multiplicity distribution in \Zg~MC events is reweighted to match the UE observed in data, following the same procedure as in Ref.~\cite{Aad:2014jgf}.
Uncorrected \Zg~MC models overestimate the charged-particle multiplicity distributions observed in data by 50\% for low-multiplicity events.
In order to estimate the relevant weight, the events in the $Z$-peak region, defined as $70~\GeV < \Mll < 105~\GeV$, are used. 
This region is expected to include a large DY component.
The correction procedure also accounts for the effect of tracks originating from pile-up and ID track reconstruction inefficiency.
The requirement of no additional tracks associated with the dilepton vertex completely removes multi-jet, \ttbar, and diboson backgrounds.

The $\Delta z_\mathrm{vtx}^\mathrm{iso}$~distribution for events with no additional tracks at the dilepton vertex is presented in Figure~\ref{fig:selection}(a).
The structure observed at small $\Delta z_\mathrm{vtx}^\mathrm{iso}$ values is due to the vertex finding algorithm, which identifies the vertex as two close vertices in high-multiplicity DY events: the two-track vertex formed from the lepton tracks and the vertex from the UE tracks.
The 3 mm~cut significantly suppresses the DY background, at the cost of a 26\% reduction in signal yield. 
The inefficiency is related to tracks and vertices originating from additional $pp$ interactions.

Contributions from the DY \ee~and \mumu~processes can be further reduced by excluding events with a dilepton invariant mass in the $Z$-peak region.
The invariant mass distribution of muon pairs for events satisfying the exclusivity veto (exactly two tracks at the dilepton vertex, $\Delta z_\mathrm{vtx}^\mathrm{iso}>3$~mm) is presented in Figure~\ref{fig:selection}(b) (where the excluded $Z$-peak region is indicated by dashed lines). 
The figure shows that the MC description of the \Mmm~distribution is satisfactory.
To further suppress the proton dissociative backgrounds, the lepton pair is required to have small total transverse momentum ($\ptll<1.5~\GeV$).
This is shown in Figure~\ref{fig:selection}(c), which displays the di-muon transverse momentum distribution for events outside the $Z$ region that satisfy the exclusivity veto.
The $\ptll$~resolution below 1.5~\GeV~is approximately 0.3~\GeV~for the electron channel and 0.2~\GeV~for the muon channel.

The result of each step of the exclusive selection applied to the data, signal and background samples is shown in Table~\ref{tab:eeCutflow}. 
After all selection criteria are applied, 869 events remain for the electron channel, and 2124 events are selected in the muon channel. 
From simulations, approximately half are expected to originate from exclusive production. 
The number of selected events in the data is below the expectation from the simulation, with an observed yield that
is approximately 80\% of the sum of simulated signal and background processes (see discussion in Section~\ref{sec:Results}).

\begin{table*}[t!]
\footnotesize
\caption{Effect of sequential selection requirements on the number of events selected in data, compared to the number of predicted signal and background events for electron and muon channels. Predictions for exclusive and single-dissociative event yields do not take into account proton absorptive corrections.}
\label{tab:eeCutflow}
\begin{center}

\resizebox{\linewidth}{!}{%
\begin{tabular}{l|cccccccc|c|c}
\hline  \hline
 & \multicolumn{3}{ c }{\gmgmll } & \Zg &  Multi-&\Zg & & Di-& Total  & \\
\cline{2-4} 
 
 Selection  & Signal & S-diss. & D-diss. & $\rightarrow \ll$ & jet & $\rightarrow \tautau$ &\ttbar& boson & predicted & Data \\

\hline
\multicolumn{11}{ c }{Electron channel ($\ell=e$)} \\
\hline
     
Preselection & 898 & 2096 & 2070 & 1\,460\,000 & 83\,000 & 3760 & 4610 & 1950 & 1\,560\,000 & 1\,572\,271 \\

Exclusivity veto  & 661 & 1480 & 470 & 3140 & 0 & 9 & 0 & 5 & 5780 & 5410\\

$Z$ region removed  & 569 & 1276 & 380 & 600  & 0 & 8 & 0 & 3 & 2840 & 2586\\

$\ptll<1.5$~\GeV  & 438 & 414 & 80 & 100 & 0 & 2 & 0 & 0 & 1030 & 869 \\

\hline
\multicolumn{11}{ c }{Muon channel ($\ell=\mu$)} \\
\hline

Preselection & 1774 & 3964 & 4390 & 2\,300\,000 & 98\,000 & 7610  & 6710 & 2870 & 2\,420\,000 & 2\,422\,745 \\

Exclusivity veto  & 1313 & 2892 & 860 & 3960 & 3 & 8 & 0 & 6 & 9040 & 7940\\

$Z$ region removed  & 1215 & 2618 & 760 & 1160  & 3 & 8 & 0 & 3 & 5760 & 4729\\

$\ptll<1.5$~\GeV  & 1174 & 1085 & 160 & 210 & 0 & 3 & 0 & 0 & 2630 & 2124\\

\hline \hline

\end{tabular}
}

\end{center}
\end{table*}

After the final exclusive event selection, there is still a significant
contamination from DY, single- and double-dissociative processes. 
Scaling factors for signal and background processes are estimated by a binned maximum-likelihood fit of the sum of the simulated distributions contained in the MC templates for the various processes, to the measured dilepton acoplanarity (\acoll) distribution. 
The fit determines two scaling factors, defined as the ratios of the number of observed to the number of expected events based on the MC predictions, for the exclusive
($R^\textrm{excl.}$) and single-dissociative ($R^\textrm{s\textnormal{-}diss.}$) templates.
The double-dissociative and DY contributions are fixed to the MC predictions in the fit procedure. 
Contributions from other background processes are found to be negligible.

Figure \ref{fig:fitYield} shows the \ee~and \mumu~acoplanarity distributions in data overlaid with the result of the fit to the shapes from MC simulations for events satisfying all selection requirements.
The results from the best fit to the data for the electron channel are:
$R_{\gmgmee}^{\mathrm{excl.}} = 0.863 \pm 0.070~(\mathrm{stat.})$
for the signal scaling factor and
$R_{\gmgmee}^{\mathrm{s\textnormal{-}diss.}} = 0.759 \pm 0.080~(\mathrm{stat.})$
for the single-dissociative scaling factor.
Similarly, for the muon channel the results are:
$R_{\gmgmmumu}^{\mathrm{excl.}} = 0.791 \pm 0.041~(\mathrm{stat.})$
and
$R_{\gmgmmumu}^{\mathrm{s\textnormal{-}diss.}} = 0.762 \pm0.049~(\mathrm{stat.}).$
The central values and statistical uncertainties on $R^\textrm{excl.}$ are strongly correlated with the central values and uncertainties on $R^{\mathrm{s\textnormal{-}diss.}}$, respectively.

\begin{figure}[t!]
\begin{center}
  \includegraphics[width=0.49\textwidth]{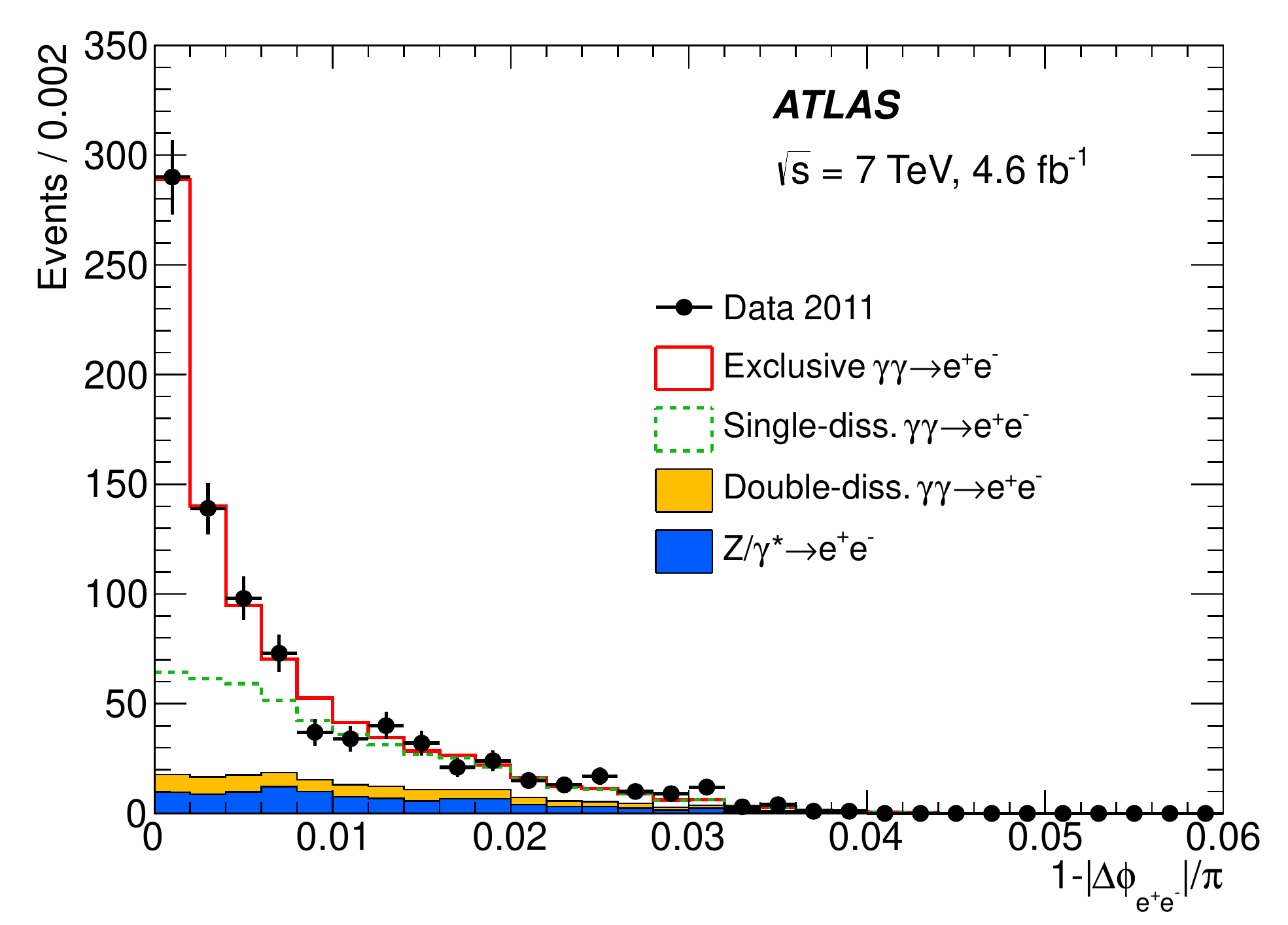}\put(-110,-8){{\footnotesize(a)}}
  %\vskip 15pt
  \includegraphics[width=0.49\textwidth]{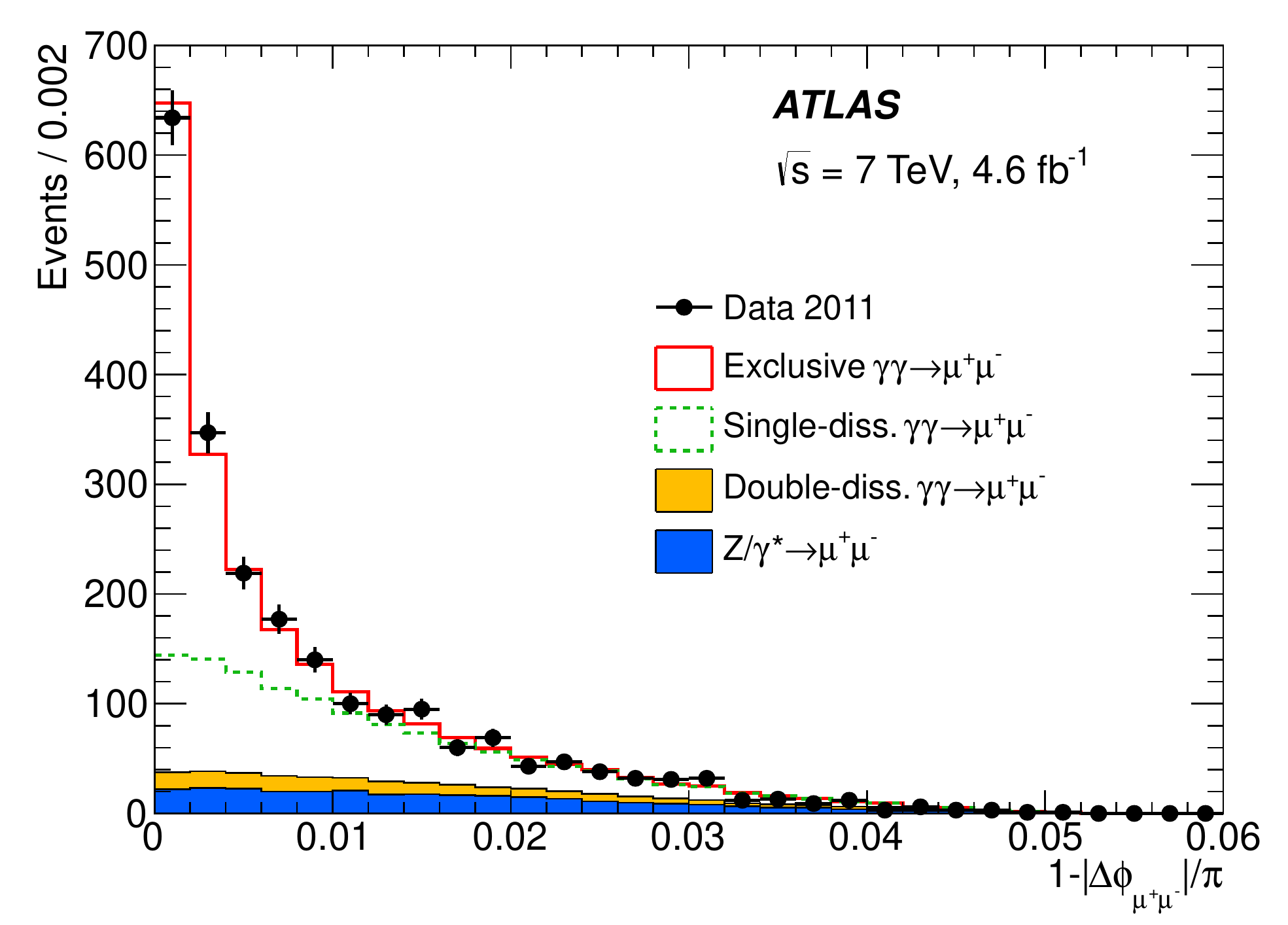}\put(-110,-8){{\footnotesize(b)}}
\end{center}
\vskip -15pt
\caption{
(a) Di-electron and (b) di-muon acoplanarity distributions for the selected sample after exclusivity requirements. Data are shown as points with statistical error bars. The stacked histograms, in top-to-bottom order, represent the simulated exclusive signal, and the single-dissociative, double-dissociative and DY backgrounds. The exclusive and single-dissociative yields are determined from the fit described in the text.
\label{fig:fitYield}
}
\end{figure}

\section{Systematic uncertainties and cross-checks}
\label{sec:sys}
The different contributions to the systematic uncertainties are described below.
The dominant sources of systematic uncertainty for both the electron and muon channels are related to background modelling.

\medskip
%\noindent{\it Lepton selection.} 
The uncertainty on the electron and muon selection includes uncertainties on the electron energy or muon momentum scale and resolution, 
as well as uncertainties on the scale factors applied to the simulation in order to reproduce the trigger,
reconstruction and identification efficiencies for electrons or muons measured in the data. 
The lepton energy or momentum scale correction uncertainties are obtained from a comparison of the
\Z~boson invariant mass distribution in data and simulation, while the uncertainties on the scale factors are
derived from a comparison of tag-and-probe results in
data and simulations~\cite{ref29, Aad:2014qja, Aad:2014rra, ref27}.
The overall effect on the exclusive \gmgmll~cross-sections is approximately 1--3\%, where the dominant electron uncertainties originate from the electron reconstruction and identification and the dominant muon uncertainty originates from the trigger.

\medskip
%\noindent{\it Backgrounds.} 
The uncertainty on the contribution of DY processes mainly accounts for disagreements between data and simulations which are related to the reweighting procedures of the charged-particle multiplicity (10\%) and \ptll~(5\%) distributions. It also includes a 5\% contribution for the PDF and scale uncertainties in modelling DY processes, as well as a 5\% statistical uncertainty on the \Zg~MC samples after event selection.
An overall normalisation uncertainty of 20\% is assigned to cover all these effects.
Because of the similar shapes of the DY and single-proton dissociative \gmgmll~components in the fitted acoplanarity distribution, this uncertainty on the DY normalisation is partly absorbed by the single-dissociative contribution.
The 20\% uncertainly has a 1.2\% effect on the exclusive cross-section for the electron channel and 1\% for the muon channel.

In order to estimate the double-proton dissociative \gmgmll~uncertainty, this contribution is varied according to the photon PDF uncertainties, defined at 68\% confidence level and evaluated using NNPDF2.3QED replicas~\cite{Ball:2013hta}.
The photon PDF are affected by sizeable uncertainties, typically of the order of 50\%. 
The resulting uncertainty on the exclusive cross-sections related to double-dissociative background uncertainty is 1.9\% for the electron channel and 1.7\% for the muon channel.

\medskip
%\noindent{\it Template shapes.} 
The uncertainty arising from the choice of acoplanarity shapes in the fit procedure is evaluated by refitting the data with different template distributions.
A small deviation of the proton elastic form-factors~\cite{Belushkin:2006qa} from the standard dipole parameterisation used in the simulations has a 0.2\% effect on the exclusive cross-sections.
This effect is estimated by reweighting the equivalent photon spectra in signal MC events to agree with the model predictions.
The impact of the shape uncertainty in the single-dissociative
template is evaluated by reweighting the corresponding MC events with an exponential modification factor $\propto \exp\left[-a (\ptll)^2\right ]$. 
A value of $a = 0.05~\GeV^{-2}$~is extracted from the data (before the $\ptll<1.5~\GeV$ selection) to improve the shape agreement with the simulation, shown in Figure~\ref{fig:selection}(c). 
Propagating these weights to the acoplanarity distribution and the
signal extraction results in a 0.9\% change of signal yields.

Possible mis-modelling of the angular resolution of the tracking detectors~\cite{Aad:2010bx} measuring the lepton tracks could also distort the shape of the signal template, and leads to uncertainties of up to 0.3\% (0.2\%) in the electron (muon) channel.

\medskip
%\noindent{\it Pile-up description.} 
The systematic effect related to the pile-up description is estimated from data-to-MC comparisons of the \pT- and $\eta$-dependent density of tracks originating from pile-up, as in Ref. \cite{Aad:2014jgf}.
The resulting uncertainty on the cross-sections is 0.5\%.

\medskip
%\noindent{\it Vertex isolation efficiency.} 
The dilepton vertex isolation efficiency is studied by comparing the spatial distribution of tracks originating from pile-up in MC simulations and in data.
The effect of mis-modelling of the vertex isolation efficiency is determined by comparing the efficiency in data and simulations for different $\Delta z_\mathrm{vtx}^\mathrm{iso}$~values (varied between 2 mm and 5 mm, where the sensitivity of the measurements to the level of background is maximal). 
The relative variations between the data and simulations are found to be at most 1.2\%, which is taken as a systematic uncertainty.

\medskip
%\noindent{\it LHC beam effects.} 
The LHC beam energy uncertainty is evaluated to be 0.7\%, following Ref.~\cite{Aad:2014jra}. 
This affects the exclusive cross-sections by 0.4\% and is considered as a systematic effect.
The impact of the non-zero crossing angles of the LHC beams at the ATLAS interaction point is estimated by applying a relevant Lorentz transformation to generator-level lepton kinematics for signal MC events.  
This results in a 0.3\% variation and is taken as a systematic uncertainty.

\medskip
%\noindent{\it QED FSR in DY \ee~events.} 
The effect of QED FSR is predicted to be small (below 1\%) in exclusive \gmgmll~reactions \cite{Denner:1998tb}.
However, as experimental corrections for electrons are derived from \Zee~and \wenu~processes including significant QED FSR effects, 
these corrections may not be directly applicable to the exclusive dilepton signal MC events without QED FSR simulation.
A possible bias in the electron efficiencies
is studied by comparing DY \ee~MC events with and without QED FSR photons being emitted.
The observed difference in the efficiency to trigger, reconstruct and identify electron pairs is 0.8\%, which is taken as a systematic uncertainty.

\medskip
Additional tests of the maximum-likelihood fit stability are performed by comparing different bin widths and fit ranges. 
Starting from the nominal number of 30 bins in the fit range $0 \leq \acoll \leq 0.06$, 
variations of the bin width ($0.002 \pm 0.001$) 
and fit range from $[0,~0.03]$ to $[0,~0.09]$
produce relative changes of at most 0.9\%.
Since these variations are strongly correlated with the statistical uncertainties, no additional systematic uncertainty is assigned in this case.

Table~\ref{tab:systSummary} summarises the contributions to the systematic uncertainty on the exclusive cross-sections from the different sources.
\begin{table*}[t!]
\footnotesize
  \caption{Summary of systematic uncertainties on the exclusive cross-section measurement for the electron and muon channels.
The data statistical uncertainties are also given for comparison.}
  \label{tab:systSummary}
  \begin{center}
    \begin{tabular}{lcc}
      \hline \hline 
       & \multicolumn{2}{c}{Uncertainty [\%]} \\
      \multicolumn{1}{ l }{Source of uncertainty} & \multicolumn{1}{c}{\gmgmee}&\multicolumn{1}{c}{\gmgmmumu} \\ 
	\hline
    
    Electron reconstruction & & \\ 
    and identification efficiency &  1.9 & -  \\
    Electron energy scale & & \\ 
    and resolution &  1.4 & -  \\
	Electron trigger efficiency & 0.7 & - \\	
	
	Muon reconstruction efficiency & - & 0.2  \\
	
	Muon momentum scale & & \\ 
	and resolution & - & 0.5 \\
	
	Muon trigger efficiency & - & 0.6 \\
	Backgrounds & 2.3 & 2.0 \\
	Template shapes & 1.0 & 0.9 \\
	Pile-up description  & 0.5 & 0.5 \\
	Vertex isolation efficiency & 1.2 & 1.2 \\
	LHC beam effects & 0.5 & 0.5 \\
	QED FSR in DY \ee & 0.8 & - \\
	Luminosity & 1.8 & 1.8 \\
	\hline
	Total systematic uncertainty & 4.3 & 3.3 \\
	\hline
	Data statistical uncertainty & 8.2 & 5.1 \\
      \hline \hline
    \end{tabular}
  \end{center}
\end{table*}
The total systematic uncertainty 
is formed by adding the individual contributions in quadrature for each analysis channel, including the uncertainty on the integrated luminosity.
Control distributions of the dilepton transverse momentum for events satisfying the selection criteria
listed in Table~\ref{tab:eeCutflow} are shown in Figure~\ref{fig:control}, with the exclusive and single-dissociative yields normalised according to the fit results. Here an additional cut on the dilepton acoplanarity ($\acoll<0.008$) is used, instead of the cut on total transverse momentum ($\ptll<1.5~\GeV$).
The MC predictions for the shapes of dilepton distributions are found to be in good agreement with the data.

\begin{figure}[!htbp]
\begin{center}
  \includegraphics[width=0.49\textwidth]{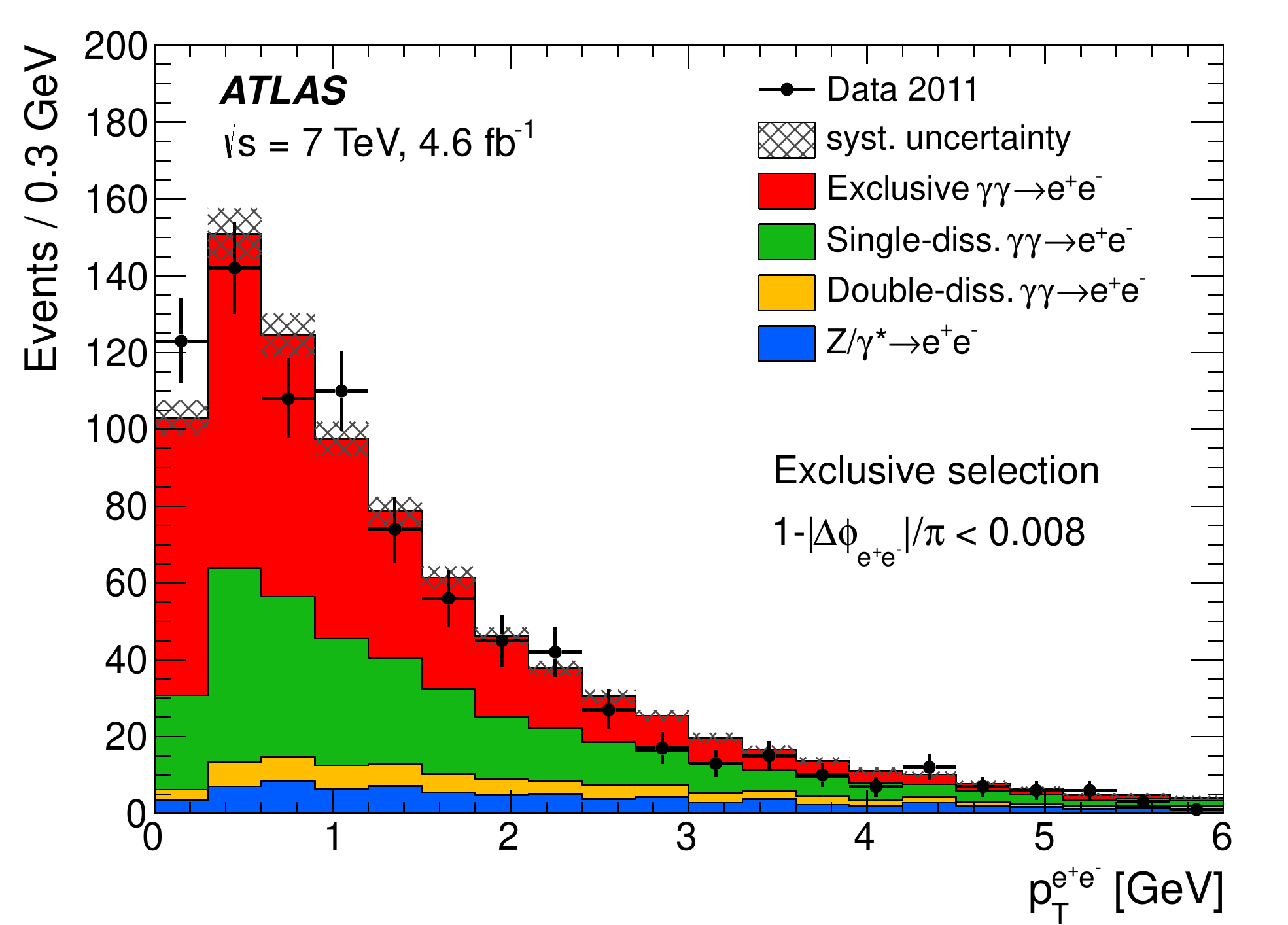}\put(-110,-8){{\footnotesize(a)}}
  \includegraphics[width=0.49\textwidth]{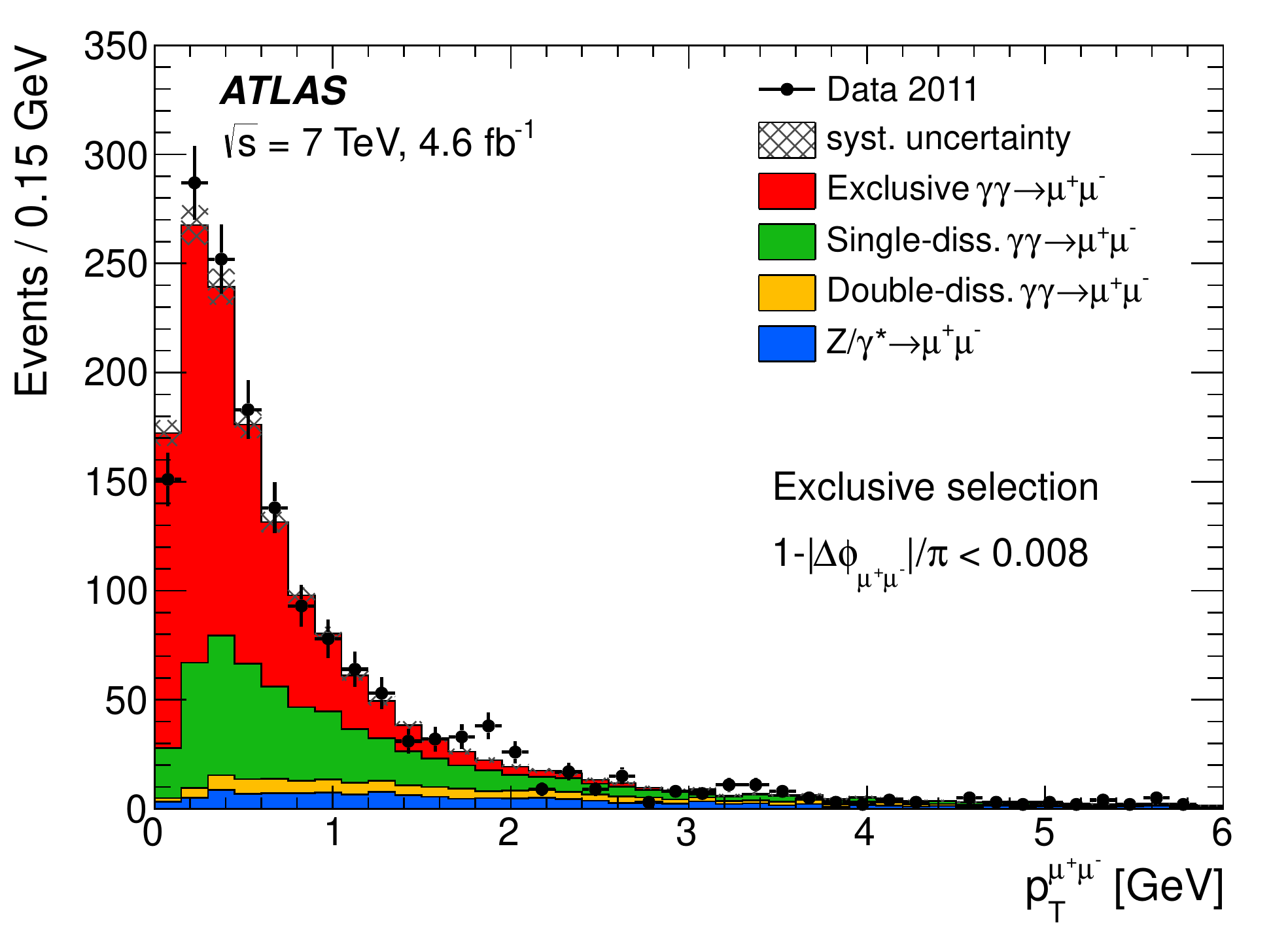} \put(-110,-8){{\footnotesize(b)}}
\end{center}
\vskip -15pt
\caption{
Control distributions of (a) the di-electron and (b) the di-muon transverse momentum for events passing the exclusivity veto together with the other selection criteria described in Section~\ref{sec:cross-sec}, and passing a cut on the dilepton acoplanarity (${1 - \left | \Delta\phi_{\leplep} \right | /\pi}<0.008$), instead of the total transverse momentum. Data are shown as points with statistical error bars, while the histograms, in top-to-bottom order, represent the simulated exclusive signal, and the single-dissociative, double-dissociative and DY backgrounds. Systematic uncertainties on the signal events are shown by the black-hashed regions. The exclusive and single-dissociative yields are determined from the fit described in the text.
\label{fig:control}
}
\end{figure}

%\newpage
\section{Results and comparison to theory}
\label{sec:Results}

The exclusive \gmgmll~cross-sections reported in this article are restricted to the fiducial regions defined in Table~\ref{tab:fidRegion}.
The event selection results in an acceptance times efficiency of 19\% for the electron channel and 32\% for the muon channel. The fiducial cross-sections are given by the product of the measured signal scale factors by the exclusive cross-sections predicted, in the fiducial region considered, by the EPA calculation:
\noindent
$$\sigma_{\gmgmll}^{\mathrm{excl.}} = R_{\gmgmll}^{\mathrm{excl.}} \cdot \sigma^{\mathrm{EPA}}_{\gmgmll} ~.$$

\par\noindent
For the \ee~channel,
\begin{eqnarray*}
R_{\gmgmee}^{\mathrm{excl.}} &= &0.863 \pm 0.070~(\mathrm{stat.}) \pm 0.037~(\mathrm{syst.}) \pm 0.015~(\mathrm{theor.})~, \\
\sigma_{\gmgmee}^{\mathrm{EPA}} &= &0.496~\pm~0.008~(\mathrm{theor.})~\pb ~.
\end{eqnarray*}
The theoretical uncertainties are fully correlated between $R_{\gmgmee}^{\mathrm{excl.}}$ and $\sigma^{\mathrm{EPA}}_{\gmgmee}$, and cancel each other in the cross-section extraction procedure. They are related to the proton elastic form-factors (1.6\%) and to the higher-order electroweak corrections~\cite{Denner:1998tb} not included in the calculations (0.7\%). The proton form-factor uncertainty is conservatively estimated by taking the full difference between the calculations using the standard dipole form-factors and the improved model parameterisation including pQCD corrections from Ref.~\cite{Belushkin:2006qa}. The latter includes a fit uncertainty and the prediction furthest away from the dipole form-factors is chosen.

\par\noindent
Similarly, for the \mumu~channel,
\begin{eqnarray*}
R_{\gmgmmumu}^{\mathrm{excl.}} &= & 0.791 \pm 0.041~(\mathrm{stat.}) \pm 0.026~(\mathrm{syst.}) \pm 0.013~(\mathrm{theor.})~, \\
\sigma_{\gmgmmumu}^{\mathrm{EPA}} & = & 0.794~\pm~0.013~(\mathrm{theor.})~\pb~.
\end{eqnarray*}

\par\noindent
The resulting fiducial cross-section for the electron channel is measured to be
\noindent
$$\sigma_{\gmgmee}^{\mathrm{excl.}} = 0.428~\pm~0.035~(\mathrm{stat.})~\pm~0.018~(\mathrm{syst.}) \ \pb~.$$
\noindent
This value can be compared to the theoretical prediction, including absorptive corrections to account for the finite size of the proton~\cite{Dyndal:2014yea}: 
\noindent
$$\sigma_{\gmgmee}^{\mathrm{EPA,~corr.}} = 0.398~\pm~0.007~(\mathrm{theor.})~\pb~.$$
\noindent
For the muon channel, the fiducial cross-section is measured to be
\noindent
$$\sigma_{\gmgmmumu}^{\mathrm{excl.}} = 0.628~\pm~0.032~(\mathrm{stat.}) ~\pm~0.021~(\mathrm{syst.}) \ \pb~,$$
\noindent
to be compared with ~\cite{Dyndal:2014yea}: 
\noindent
$$\sigma_{\gmgmmumu}^{\mathrm{EPA,~corr.}} = 0.638~\pm~0.011~(\mathrm{theor.})~\pb~.$$ 
The uncertainty of each prediction includes an additional 0.8\% uncertainty related to the modelling of proton absorptive corrections.
It is evaluated by varying the effective transverse size of the proton by 3\%, according to Ref.~\cite{Aaron:2007ab}.
Figure~\ref{fig:ratioemu} shows the ratios of the measured cross-sections to the EPA calculations and to the prediction with the inclusion of absorptive corrections.
The measurements are in agreement with the predicted values corrected for proton absorptive effects.
The figure includes a similar CMS cross-section measurement~\cite{Chatrchyan:2011ci}.

\begin{table*}[t!]
\footnotesize
  \caption{Definition of the electron and muon channel fiducial regions for which the exclusive cross-sections are evaluated.}
  \label{tab:fidRegion}
  \begin{center}
    \begin{tabular}{lcc}
      \hline \hline 
	Variable  & Electron channel & Muon channel \\ \hline
    $\pT^\ell$ & $>12$~\GeV & $>10$~\GeV \\
    $|\eta^\ell|$ & $<2.4$ & $<2.4$ \\
    $\Mll$ & $>24$~\GeV & $>20$~\GeV \\
      \hline \hline
    \end{tabular}
  \end{center}
\end{table*}

\begin{figure}[!htbp]
\centering
\includegraphics[width=0.49\textwidth]{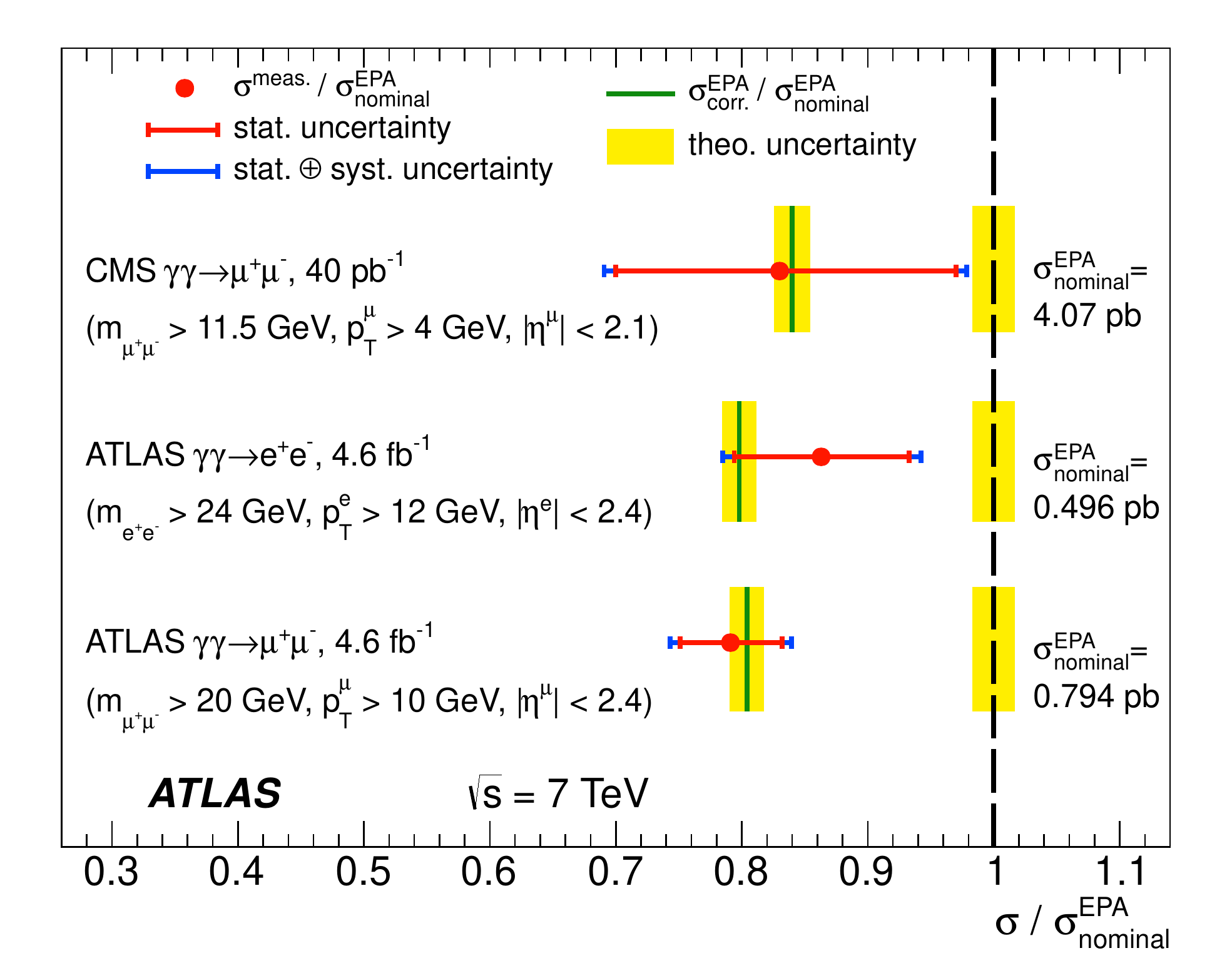}
\caption{
Comparison of the ratios of measured (red points) and predicted (solid green lines) cross-sections to the uncorrected EPA calculations (black dashed line). Results for the muon and electron channels are also compared with a similar CMS measurement~\cite{Chatrchyan:2011ci}. The inner red error bar represents the statistical error, and the blue bar represents the total error on each measurement.
The yellow band represents the theoretical uncertainty of 1.8\% (1.7\%) on the predicted (uncorrected EPA) cross-sections, assumed to be uniform in the phase space of the measurements.
\label{fig:ratioemu}
}
\end{figure}

\section{Conclusion}
\label{sec:summary}
Using $4.6~\ifb$ of data from $pp$ collisions at a centre-of-mass energy of $7~\TeV$ the fiducial cross-sections for exclusive \gmgmll ($\ell=e,~\mu$) reactions have been measured with the ATLAS detector at the LHC.
Comparisons are made to the theory predictions based on EPA calculations, as included in the \herwigpp~MC generator.
The corresponding data-to-EPA signal ratios for the electron and muon channels are consistent with the recent CMS measurement and indicate a suppression of the exclusive production mechanism in data with respect to EPA prediction. 
The observed cross-sections are about 20\% below the nominal EPA prediction, and consistent with the suppression expected due to proton absorption contributions.
The MC predictions for the shapes of the dilepton kinematic distributions, including both the exclusive signal and the background dominated by two-photon production of lepton pairs with single-proton dissociation, are also found to be in good agreement with the data.
With its improved statistical precision compared to previous measurements, this analysis provides a better understanding of the physics of two-photon interactions at hadron colliders.

\section*{Acknowledgements}
% Acknowledgements for papers with collision data
% Version 23-Mar-2015

% Standard acknowledgements start here
%----------------------------------------------
We thank CERN for the very successful operation of the LHC, as well as the
support staff from our institutions without whom ATLAS could not be
operated efficiently.

We acknowledge the support of ANPCyT, Argentina; YerPhI, Armenia; ARC,
Australia; BMWFW and FWF, Austria; ANAS, Azerbaijan; SSTC, Belarus; CNPq and FAPESP,
Brazil; NSERC, NRC and CFI, Canada; CERN; CONICYT, Chile; CAS, MOST and NSFC,
China; COLCIENCIAS, Colombia; MSMT CR, MPO CR and VSC CR, Czech Republic;
DNRF, DNSRC and Lundbeck Foundation, Denmark; EPLANET, ERC and NSRF, European Union;
IN2P3-CNRS, CEA-DSM/IRFU, France; GNSF, Georgia; BMBF, DFG, HGF, MPG and AvH
Foundation, Germany; GSRT and NSRF, Greece; RGC, Hong Kong SAR, China; ISF, MINERVA, GIF, I-CORE and Benoziyo Center, Israel; INFN, Italy; MEXT and JSPS, Japan; CNRST, Morocco; FOM and NWO, Netherlands; BRF and RCN, Norway; MNiSW and NCN, Poland; GRICES and FCT, Portugal; MNE/IFA, Romania; MES of Russia and NRC KI, Russian Federation; JINR; MSTD,
Serbia; MSSR, Slovakia; ARRS and MIZ\v{S}, Slovenia; DST/NRF, South Africa;
MINECO, Spain; SRC and Wallenberg Foundation, Sweden; SER, SNSF and Cantons of
Bern and Geneva, Switzerland; NSC, Taiwan; TAEK, Turkey; STFC, the Royal
Society and Leverhulme Trust, United Kingdom; DOE and NSF, United States of
America.

The crucial computing support from all WLCG partners is acknowledged
gratefully, in particular from CERN and the ATLAS Tier-1 facilities at
TRIUMF (Canada), NDGF (Denmark, Norway, Sweden), CC-IN2P3 (France),
KIT/GridKA (Germany), INFN-CNAF (Italy), NL-T1 (Netherlands), PIC (Spain),
ASGC (Taiwan), RAL (UK) and BNL (USA) and in the Tier-2 facilities
worldwide.
%----------------------------------------------

%-------------------------------------------------------------------------------
% If you use biblatex and either biber or bibtex to process the bibliography 
% just say \printbibliography here
\printbibliography
% If you want to use the traditional BibTeX you need to use the syntax below.
%\bibliographystyle{bibtex/bst/atlasBibStyleWoTitle}
%\bibliography{excdilept_paper,bibtex/bib/atlas-paper}
%-------------------------------------------------------------------------------

%-------------------------------------------------------------------------------
% Print the list of contributors to the analysis
% The argument gives the fraction of the text width used for the names
%-------------------------------------------------------------------------------
%\clearpage
%\PrintAtlasContribute{0.30}

%-------------------------------------------------------------------------------
%\clearpage
%\appendix
%\part*{Auxiliary material}
%\input{Sec10_Appendix}

\newpage 
% ATLAS Collaboration author list
% Data extracted on 11-Aug-2015 for paper reference STDM-2014-13
\begin{flushleft}
{\Large The ATLAS Collaboration}

\bigskip

G.~Aad$^{\rm 85}$,
B.~Abbott$^{\rm 113}$,
J.~Abdallah$^{\rm 151}$,
O.~Abdinov$^{\rm 11}$,
R.~Aben$^{\rm 107}$,
M.~Abolins$^{\rm 90}$,
O.S.~AbouZeid$^{\rm 158}$,
H.~Abramowicz$^{\rm 153}$,
H.~Abreu$^{\rm 152}$,
R.~Abreu$^{\rm 116}$
% ...
\bigskip
\\
% ...
$^{11}$ Institute of Physics, Azerbaijan Academy of Sciences, Baku, Azerbaijan\\
% ...
$^{85}$ CPPM, Aix-Marseille Universit{\'e} and CNRS/IN2P3, Marseille, France\\
% ...
$^{90}$ Department of Physics and Astronomy, Michigan State University, East Lansing MI, United States of America\\
% ...
$^{107}$ Nikhef National Institute for Subatomic Physics and University of Amsterdam, Amsterdam, Netherlands\\
% ...
$^{113}$ Homer L. Dodge Department of Physics and Astronomy, University of Oklahoma, Norman OK, United States of America\\
% ...
$^{116}$ Center for High Energy Physics, University of Oregon, Eugene OR, United States of America\\
% ...
$^{151}$ Institute of Physics, Academia Sinica, Taipei, Taiwan\\
$^{152}$ Department of Physics, Technion: Israel Institute of Technology, Haifa, Israel\\
$^{153}$ Raymond and Beverly Sackler School of Physics and Astronomy, Tel Aviv University, Tel Aviv, Israel\\
% ...
$^{158}$ Department of Physics, University of Toronto, Toronto ON, Canada\\
% ...
\end{flushleft}

%\addcontentsline{toc}{part}{Auxiliary material}

%-------------------------------------------------------------------------------

\end{document}